\def\pc{\,{\rm pc}}
\def\pc2{\,{{\rm pc}^{-2}}}
\def\kpc{\,{\rm kpc}}

\def\cmm2{{\,\rm cm^{-2}}}
\def\cm2{{\,{\rm cm}^2}}
\def\cmm3{{\,{\rm cm}^{-3}}}
\def\gcmm3{{\,{\rm g\,cm^{-3}}}}

\def\Msol{{M_\odot}}
\def\deg{{^\circ}}

\def\BE{\begin{equation}}
\def\EE{\end{equation}}
\documentstyle[rotate,epsf,12pt]{article}
\begin{document}
\baselineskip=24pt
\pagestyle{empty}
\bigskip
\begin{center}
\bigskip
{\Large \bf Gravitational Microlensing and the Structure of the
Inner Milky Way}

\bigskip

{G. Gyuk}

\bigskip

{\normalsize\rm Department of Physics, Enrico Fermi Institute, University of Chicago, 5630 Ellis Avenue,
Chicago IL 60637-1433}

\medskip

{\normalsize\rm NASA/Fermilab Astrophysics Center, Fermi National Accelerator
Laboratory, Batavia, IL 60510-0500}

\end{center}

\centerline{\bf ABSTRACT}
\bigskip

We analyze the first-year MACHO collaboration observations of microlensing
towards the Galactic center using a new direct likelihood technique that
is sensitive to the distribution of the events on the sky. We consider the
full set of 41 events, and calculate the direct likelihood against a
simply-parameterized Galactic model consisting of either a gaussian or 
exponential bar and a double exponential disk. Optical depth maps are
calculated taking into account the contribution of both disk lenses and
sources. We show that based on the presently available data, a slope in
the optical depth has been clearly detected ($3\sigma$) in Galactic
latitude and that there are indications of a small slope in Galactic
longitude.  We discuss limits that can be set on the mass, angle and axis
ratio of the Galactic bulge. We show that based on microlensing
considerations alone, $M_{Bulge}>1.5\times 10^{10}\Msol$ at the 90\%
confidence level and that the bulge inclination angle is less than $30
\deg$ also at the 90\% confidence level. The mostly likely bar mass is
$M_{Bulge}=3.5\times10^{10}\Msol$. Such a high mass would imply a low
MACHO fraction for the halo. We consider disk parameters and
show that there are two degeneracies between the effects of a disk and
those of a bar on the optical depths. Finally, we discuss how to break
these degeneracies and consider various strategies for future microlensing
observations.

\newpage
\pagestyle{plain}
\setcounter{page}{1}
\newpage

\section{Introduction}

Although the Galaxy has been studied for a long time, determining its
structure has proved to be an extremely difficult task. Our basic picture
of the Milky Way as a spiral galaxy with a roughly exponentially falling
disk, a central bulge and an extended halo has been settled for a
considerable time\cite{galstudy}. Unfortunately, despite our 
wealth of detailed knowledge of the Galaxy, gaining a more precise
knowledge of its global parameters, such as we have for external galaxies,
is complicated by our position. Even such basic quantities as the scale
length of the disk or the rotation curve of our own Galaxy are less well
known than those of many external galaxies, due to the unfavorable
geometry and intervening dust and gas\cite{rotation,scalelength}. Because
of our privileged position within it, the Milky Way provides us with a
unique opportunity for studying questions such as how galaxies form or
what their major constituents are. Ironically, many of the techniques we
have for studying such questions are essentially measurements of light,
while the basic questions have more to do with the mass and its
distribution. The translation between these measurements and the
information we would like is complicated by our fundamental lack of knowledge of the
stellar mass function for small masses.

Gravitational microlensing searches are particularly exciting because,
unlike other astrophysical observations, they can detect objects
regardless of their luminosity. Within its mass range, a microlensing
search is sensitive to the integrated density in massive compact objects
of any type between the observer and source star. Originally intended for
probing the baryonic component of the Galactic halo, gravitational
microlensing has been increasingly recognized as a powerful new way to
probe the structure of our entire Galaxy\cite{Galprobe}. Although
microlensing can also be used to probe the stellar mass function, we will
concentrate on using it to directly constrain the mass density.

Over the past few years, microlensing has moved rapidly from a proposal to
an established fact. With its characteristically shaped light curve and
achromaticity, there can be little doubt that microlensing has been
observed in abundance towards the galactic bulge.  Over one hundred events
have now been observed by three collaborations, MACHO, DUO and OGLE, in
the general direction of Baade's Window \cite{MACHO1st,oglebulge,DUO}.
Small theoretically expected modifications to the light curve, such as
effects from parallax and binary lenses have also been observed in some
events, further confirming their interpretation as
microlensing\cite{lightcurve}.

	One of the most exciting of the recent microlensing results has
been the observation of many more microlensing events in the direction of
the Galactic bulge than had been predicted. One explanation for the higher
than expected event rates is that the the Milky-Way is actually a barred
spiral, with the bar oriented almost directly towards us\cite{barsug}. A
bar pointing at us would concentrate mass along our line of sight,
increasing the number of microlensing events, but not leave an obvious
signature of asymmetry in other observations.  The suggestion that the
galactic bulge is actually a bar is not new. As far back as 1964 de
Vaucouleurs suggested a bar as a possible explanation for similarities
between the gas dynamics seen towards the Galactic center and that seen in
barred galaxies\cite{deVauc}. This idea, however, was not universally
embraced. More recently, non-circular motions of the gas towards the
Galactic center have been accounted for by a bar. A variety of other
observations, such as star counts and luminosity studies, also indicate
such a structure\cite{Binney,whitelock,barevid}. Consistent with this,
although independently not compelling, are the infrared maps from the
Diffuse Infra-Red Background Experiment (DIRBE) on the COBE satellite
\cite{Weiland,Dwek}. Taken together, these more recent observations have
led to a resurgence of the bar model. The high microlensing rate observed
towards the Galactic center is not only consistent with this new picture,
but can also be used as a probe to refine our knowledge of the Galactic
structure and in particular the Galactic Bulge. A better understanding of
the Galactic Bulge is important not only for its own sake, but also
because of its interactions with the disk and halo. A bar may be
intimately tied to the spiral density waves in our disk and as we discuss
later, its mass plays an important role in constraining our knowledge of
the baryonic content of the halo\cite{long}.

Previous work comparing the results of microlensing searches to Galactic
models has mostly focused on a single number: the microlensing optical
depth in the direction of Baade's window\cite{earlymap}. However, in this region
the optical depth is expected to be a rapidly varying function of
latitude. The use of an average quantity, such as reported by the MACHO
collaboration, limits our ability to match predictions from Galactic
models with the data.  Further, because a given optical depth is
achievable in a variety of ways we cannot discriminate between the many
possible models on the basis of this single number. It is only with using
the gradient information, by a comparison of the distribution of the event
locations to a map of the microlensing optical depth, that the various
models can be sorted out. Thus we have developed a method for calculating
the relative likelihood that a given map of the optical depth would
produce the observed events. Using this likelihood technique we explore
the constraints that the MACHO collaboration first year events impose on
Galactic structure.  We consider a class of models with bars based on the
G2 and E2 models of Dwek et al. \cite{Dwek}, and a simple double
exponential disk. We calculate maps of the microlensing optical depth for
each of these models and compare them to the observed events, calculating
the likelihood as a function of the model parameters.

The paper is structured as follows: in the second section we briefly
review the phenomenon of microlensing and the optical depth for
microlensing, and develop the formalism we use in our likelihood
analysis. We then discuss the models of Galactic structure we adopt
for this paper and what constraints exist on the parameters for these
models. In section three we examine the set of events and fields that we
will use and discuss the observational efficiencies and sky coverage. We
continue with a look at the structure of the data, determine whether the
data support a gradient in the optical depth and produce a crude map of
the optical depth. Section four contains the results of applying the
likelihood formalism to the models mentioned earlier. We discuss the
limits that can be placed on various parameters and which parameters are
correlated. The fifth section discusses future directions and examines
strategies for maximizing what we learn from our microlensing
investment. Finally in the last section we summarize the main results of
this investigation.

\section{Methods}
\subsection{Microlensing Optical Depth}
	When a massive object passes by the line of sight to a distant
object, the object's image is distorted according to general relativity. Stunning
confirmations of this have been seen in a variety of systems involving
galaxy clusters and background quasars or galaxies where typical
deflection angles are on the order of an arcsecond\cite{biglenses}. In microlensing the
source is typically a distant star and the lens an intervening massive
object. With the masses and distances much smaller the deflections are on
the order of milliarcseconds, far too small to be resolved\cite{compendium}. The distortion
of the source shape is, however, not the only effect. The intervening mass also
acts as a lens, concentrating the light of the source. The amplification is
given by, 
\BE A = \frac{u^2+2}{u\sqrt{u^2+4}}\EE
\BE r_E= \sqrt{\frac{4GM d(D-d)}{c^2D}} \EE
where $r_E$ is the
Einstein ring radius, 
$u$ is the dimensionless impact parameter, $r/r_E$, $M$ is
the mass of the lens, $D$ is the distance to the source and $d$ is the
distance to the lens\cite{pac}. The amplification is more easily
measured. When the dimensionless impact parameter, $u$, is less than one,
we say that the lens is within the microlensing tube of the source. In this
regime, the amplification is greater than about 1.34, corresponding
roughly to experimental cutoffs.

	What is probability that a given star is being microlensed? If this
chance is small, as it is in all cases we consider, then it can be
expressed as the typical number of lenses in the microlensing tube.  Thus
we have the optical depth\cite{halocont}, 
\BE\tau = \int_0^L \frac{\rho}{M} \pi r_E^2 dl = \frac{4\pi G}{c^2}\int_0^D
\rho(d) \frac{d(D-d)}{D}dd,\EE 
where $\rho$ is the spatial mass density of lenses. Note that the dependence on the mass of the lens cancels out. This,
together with the lack of a lens velocity dependence (which enters into
calculations of the rate of microlensing events) is one of the great advantages of using the
microlensing optical depth. It means we need not consider the mass
spectrum of the lenses or their distribution in velocity space and can
consider only their spatial mass density. We must, however, give up the
possibility of using the distribution of event durations to learn about
the stellar mass function or velocity distribution.  In the formalism
developed in the following we consider only the optical depth.
  
	For a field with all sources at the same distance, such as the
Large Magellanic Cloud or the Small Magellanic Cloud, Eq.(3) can be used
directly to calculate the optical depth. If the source stars are at
varying distances from the observer they will sample different optical
depths. The observed depth is given by integrating over the source density
along the line of sight:
\BE 
\tau = {4 \pi G\over c^2} {\int^\infty_0 dD D^\alpha \rho_s (D)
\int^D_0 dx \rho_l (x) {x(D-x)/ D} \over \int^\infty_0 dD D^\alpha \rho_s
(D)}, 
\EE 
where $\rho_s$ is the mass density in source stars, $\rho_l$ is
the mass density in lenses, and $\alpha$ controls the source integration
volume. Two factors enter into the determination of
$\alpha$. On the one hand, as the distance from the observer increases,
the number of stars seen in a given solid angle will increase as the
square of the distance. On the other hand, as the distance from the
observer increases, fewer stars will be above the magnitude limit to be
seen by the observer. For source stars on the main sequence these effects
almost cancel out and $\alpha=0$ is appropriate. For giants, such as the
red clump giants used in the MACHO collaboration's analysis of a subset of
their data, which can be seen throughout the bulge, $\alpha=2$ is more
reasonable. Since we use the full sample, which is
composed of mainly main-sequence stars, in our analysis we use $\alpha=0$.
 
\subsection{Likelihood}
	Although the average optical depth towards Baade's window is
useful, it is not the entire story. In the standard technique one estimates
the optical depth for a sample of events by summing up the total durations
for stars that have been microlensed, $\hat{t}_i$, (weighted by the efficiency), and dividing by the
total Exposure, $E$, (star years observed):\footnote
{\baselineskip=10pt
\vtop{ \hbox{\hbox to -1pt{} Actually, the
quantities extracted from the data, $\hat{t}_i$ measure the Einstein ring
crossing time,}
\hbox{not the time that the event is magnified above
threshold. We adjust for this by using the}
\hbox{average duration for an event with
a given $\hat{t}$, $\pi \hat{t}/4$.}}}
%{Actually, the
%quantities extracted from the data, $\hat{t}_i$ measure the Einstein ring
%crossing time, not the time that the event is magnified above
%threshold. We adjust for this by using the average duration for an event with
%a given ${t}$, $\pi {t}/4$.}
\BE
\tau= \frac{\pi}{4E}\sum\frac{\hat{t}_i}{\epsilon_i}
\EE
This procedure inevitably loses information, because it
averages over the events, ignoring their locations. For regions such as the
MACHO fields, where there are strong variations of the optical depth, it is
unclear how meaningful a procedure this is. The location at which one has
measured the optical depth is undefined. This makes it difficult to
compare ones' results to models of the optical depth. Additionally, it is
difficult to reliably quantify the gradient information, even using
latitude cuts, because of the uncertainty in the ``distance'' between the
two sub-regions.

We would like to extract the gradient information and avoid the question
of precisely ``where'' the optical depth is measured. For this we must deal
with maps of the optical depth as a function of Galactic longitude and
latitude and not simply average values.  Central to this is an ability to
quantify how well a given theoretical map of the optical depth compares to
the observed events. Thus we construct a likelihood function sensitive not
only to the number and durations of the events but also their positions.
Constructing the required likelihood is not entirely trivial. It is
instructive to look first at the case for maps of the microlensing
rate. We then show how this is modified when dealing with the optical
depth.

For any small patch of the sky, $dxdy$, the expected number of events in a
time $T$ is
$T\Gamma dx dy$, where $\Gamma$ is the rate per area. If $n_i$ is the number of events actually observed in the
$i$th sky-patch then by simple Poisson statistics the likelihood of the true
rate being
$\Gamma(x,y)$, given the observations, is 
\BE
\prod_i e^{-T\Gamma dx dy} (T \Gamma dx dy)^{n_i}/{n_i!}.
\EE 
Since for a sufficiently small patch size $n_i$ will always be either 0 or
1 we arrive at
\BE
L=\exp\left({-T\int\Gamma dx dy}\right) dx^N dy^N T^N \prod_{events} \Gamma(x,y),
\EE
where $N$ is the total number of events. Dropping the $dx^N dy^N T^N$ factor, we have the relative likelihood for a model $\Gamma(x,y)$,
\BE 
L=\exp\left({-T\int\Gamma dx dy}\right) \prod_i^{events} \Gamma(x_i,y_i).
\EE 

	The case of optical depth is more subtle but parallel. We consider
a patch $dx dy dt$, where $t$ is the time coordinate. Since the optical depth
gives the probability that any given star will be lensed at a given time,
the probability of
observing $n_i$ microlensing events in progress in this patch is 
\BE
 e^{-\sigma\tau dxdy} \frac{(\sigma\tau dxdy)^{n_i}}{n_i!}. 
\EE
Here $\sigma$ is the number density of source stars observed on the sky
and $\tau$ is the optical depth in the patch. Note that this is
independent of $dt$. For the entire region then, our likelihood is simply
\BE
L=\prod^{dx dy dt}_{all~cells} e^{-\sigma\tau dxdy} \frac{(\sigma\tau
dxdy)^{n_i}}{n_i!}.
\EE 
As before, for $dx dy$ small enough, $n_i\rightarrow 0,1$
and so
\BE
 L=\prod^{dt} \exp\left({-\int \sigma\tau dxdy}\right) \prod^{dx dy dt}_{events}\sigma\tau
dxdy
\EE
where the second product is over all the cells containing events. Since
the first term is independent of time the product over all intervals $dt$ is
easy. To do the second product we note that for each event the average
time spent in the microlensing tube for an event with the measured $\hat{t}$ is
$\frac{\pi}{4}\hat{t}$. Hence we have
\BE
 L = \exp\left({-\frac{T}{dt}\int \sigma\tau dxdy}\right) \prod_{events}
(\sigma\tau(x_i,y_i)dxdy)^\frac{\pi \hat{t}_i}{4 dt}.
\EE
If all of the events have not been detected, that is the efficiency
$\epsilon < 1$, then we will be missing terms in the product over events. To
account for this we write
\BE
 L = \exp\left({-\frac{T}{dt}\int \sigma\tau dxdy}\right) \prod_{events}
(\sigma\tau(x_i,y_i)dxdy)^\frac{\pi \hat{t}_i}{4 dt \epsilon_i}
\EE
where $\epsilon_i$ is the efficiency for detecting events of length
$\hat{t}_i$. Efficiencies for present experiments are typically of order
0.5. There is an immediate difficulty with this equation however: as
$dt\rightarrow 0$ we have infinite exponents. This comes about because we
are multiplying an infinite number of finite probabilities: one for each
timeslice $dt$. What this procedure ignores is that there is a
characteristic time interval over which microlensing in a cell $dxdy$ will
be correlated. Thus we rescale the infinities with an ad
hoc correlation constant $t_0$ which we will determine later. Thus we have finally,
\BE
 L = \exp\left({-\frac{T}{t_0}\int \sigma\tau dxdy}\right) \prod_{events}
(\sigma\tau(x_i,y_i))^\frac{\pi \hat{t}_i}{4 t_0 \epsilon_i}
\EE
where we have dropped the $dx dy$ in the product.

In the limiting case where $\tau=\tau_0$, a constant over the region of
interest, we expect to recover the standard formula for optical depth. In
this case our likelihood is
\BE
 L = \exp\left({-\frac{T}{t_0}A\sigma\tau_0}\right) (\sigma\tau_0)^{\frac{\pi}{4
t_0}\sum_i \frac{\hat{t}_i}{\epsilon_i}}
\EE
where A is the area of the region and the sum is over the events observed.
 Setting the derivative with respect to $\tau_0$ equal to zero we find
\BE
 -\frac{T}{t_0} A\sigma L + \frac{\pi}{4 t_0}\sum_i
\frac{\hat{t}_i}{\epsilon_i} \frac{L}{\tau_0}=0.
\EE
So the maximum is at
\BE
\tau_0 = \frac{\pi}{4 T A \sigma} \sum_i
\frac{\hat{t}_i}{\epsilon_i} = \frac{\pi}{4 E} \sum_i
\frac{\hat{t}_i}{\epsilon_i}
\EE
which is the standard formula. Looking at Eq. 14 we
see that it can be rewritten as a scaled Poisson distribution
\BE
 L = \exp\left({-\frac{T}{t_0}A\sigma\tau_0}\right)
\left(\frac{T}{t_0}A\sigma\tau_0\right)^{\frac{\pi}{4 t_0}\sum_i
\frac{\hat{t}_i}{\epsilon_i}} \left(\frac{T}{t_0}A\right)^{-\frac{\pi}{4
t_0}\sum_i \frac{\hat{t}_i}{\epsilon_i}}
\EE
with maximum given above. This form allows us to fix $t_0$. We see that
the distribution above has a width
\BE
\left(\frac{\delta\tau}{\tau}\right)^2 = \left({\frac{\pi}{4 t_0}\sum_i
\frac{\hat{t}_i}{\epsilon_i}}\right)^{-1}.
\EE
We compare this to the expected uncertainty as calculated by Han \& Gould
\cite{HanGould}. We see that our
\BE
\frac{1}{\frac{\pi}{4 t_0}\sum_i
\frac{\hat{t}_i}{\epsilon_i}} = \frac{1}{N}
\frac{\left<P^2\right>}{\left<P\right>^2}
\EE
in their notation. This works out to 
\BE
 t_0= \frac{\pi}{4}\frac{\sum_i(\hat{t}_i/\epsilon_i)^2}{\sum_i
\hat{t}_i/\epsilon_i} 
\EE
which completely specifies our likelihood function. 

The beauty of this likelihood function approach is in its flexibility. We
can use it to directly compare the models to the data without first
calculating an ``observed'' optical depth, or we can use it to explore
what the data say about the optical depth. We will discuss later our
construction of a primitive map of optical depth from the presently
observed data. This approach also shines in the analysis of observations
of a disparate collection of lines of sight, especially if some have low
optical depth and produce no events. One simply uses a density
function which is non-simply connected and the null data is taken into
account automatically. Overlapping fields are also easy
to handle in this formalism. Areas in which one has an overlap are simply
counted as having twice the density.

We use our likelihood to rank the models we will discuss in the next
section. It is important to note that we will be able to compute only
relative likelihoods. The true structure of the Galaxy almost certainly
does not fall exactly into the classes of models we consider. By
considering a range of models that have passed a variety of other tests we
hope to include at least some models that approximate reality.

\subsection{Models}
Our Galaxy can be described loosely as consisting of three parts: a disk,
a dark halo, and a central bulge. In the following section, we describe
the models we use in our calculations of optical depths. For each
component, we indicate a range for its parameters that we feel is
reasonable based on the literature. Because the values of these parameters
are so uncertain we felt that it would indicate a false level of certainty
to use a Gaussian prior for the parameters of our models. Accordingly, we
have taken flat priors over the parameter ranges indicated. Since we
expect the amount of microlensing due to objects in the halo to be
negligible \cite{halocont} and almost constant over the small range of
directions examined, we do not model the halo. 

The Galactic density enters the equations for the optical depth in two
distinct ways: source and lens. One must therefore be very careful to keep
the two roles separate. If both the bulge and the disk have the same lens
to source ratio (or equivalently the same mass to light ratio), then the
distinction becomes meaningless and can be dropped. It is not clear that
this is the case. We handle the issue in the following manner. The visible
content of the disk, that which could be seen as sources, is relatively
well known. On the other hand, the ratio of bulge to disk source stars in
the MACHO fields has been estimated by the MACHO collaboration as around
20\% based on the fraction of the 2.2 micron flux contributed by the
disk\cite{MACHO1st}. We therefore model the disk with a source disk and a lens
disk, and adjust the bulge source fraction to give approximately an 80\%
contribution along our line of sight.

\subsubsection{Bulge}
Models of the Galactic bulge are very uncertain. The difficulties are both
practical (most fields towards the bulge suffer extremely high extinction)
and theoretical (it is extremely difficult to invert gas and stellar
dynamics to obtain potentials). Nevertheless, a number of models have
emerged as standards\cite{Dwek}. These models have widely ranging functional
forms. We consider two representative forms, Dwek et al's G2 and E2. Their
densities can be expressed
\begin{eqnarray}
~~~~~\rho_{G2} & = & 1.2172\frac{M_{BAR}}{8\pi x_0y_0z_0} e^{-{r_s^2}/{2}} \\
~~~~~\rho_{E2} & = & \frac{M_{BAR}}{8\pi x_0 y_0 z_0} e^{-r}
\end{eqnarray}
\begin{eqnarray}
~~~~~r_s & = & {\left\{\left[\left(\frac{x}{x_0}\right)^2+\left(\frac{y}{y_0}\right)^2\right]^2+\left(\frac{z}{z_0}\right)^4\right\}^{1/4}}\nonumber\\
~~~~~r & = & {\left\{\left(\frac{x}{x_0}\right)^2+\left(\frac{y}{y_0}\right)^2+\left(\frac{z}{z_0}\right)^2\right\}^{1/2}}\nonumber
\end{eqnarray}
with the major axis inclined towards us at an angle $\theta_B$, which we
will take to be between $0\deg$ and $45\deg$, with the
near side in the first Galactic quadrant. The G2 model is the best fit to
the DIRBE infrared maps \cite{Dwek}, while the E2 is favored by an analysis of
the distribution of red clump giants in the OGLE fields\cite{OGLEfields}. At the
location of the MACHO fields we will be analyzing, G2 models tend to have
high optical depths while E2 models produce optical depth considerably
less efficiently \cite{lateZhao}. By considering both models with a wide range of
parameters we hope to cover a large portion of possible bulge models.

Estimates of bulge masses cover a wide range. More recently, however,
estimates have fallen approximately in the range $(2.0 -
3.0)\times10^{10}\Msol$\cite{lateZhao}. To be conservative we consider the range $1.0 -
4.0\times10^{10}\Msol$. The bulge scale factors are less well
known. The best known of these are the two vertical scale heights. We
follow the results of Stanek et al.\cite{OGLEfields} and fix these at $0.43\kpc$ for the G2 models and
$0.25\kpc$ for the E2 models. We consider the ranges $0.3 - 2.7\kpc$ (G2) and
$0.2 - 1.8\kpc$ (E2) for both $x_0$ and $y_0$.
Following the review paper
by Kerr \& Lynden-Bell \cite{R0} we fix the distance to the bulge to be
the IAU recommended value of 8.5kpc.

\subsubsection{Disk}
As discussed above the structure of the luminous component of the disk is
relatively better known than a possible dark component \cite{morelumdisk}. Thus we
consider two separate disks. The first is a thin luminous double
exponential disk,
\begin{equation}
\rho_{lum} = \frac{\Sigma_{lum}}{2 r_z}\exp\left({-\frac{z}{r_z}}\right)\exp\left({\frac{r_0-r}{r_d}}\right), 
\end{equation}
composed of luminous stars that can serve as sources. The parameters for
this disk are fixed: $\Sigma_{lum}=15\Msol \pc2$, $r_z=0.3 \kpc$, and
$r_d=3.5\kpc$.\cite{GWK,morelumdisk} Not all stars in the disk are bright enough to been
seen, however, and in fact there is evidence that the disk contains
considerable mass beyond that which is visible, perhaps distributed
somewhat differently from the luminous mass \cite{sigma0,sigma1,sigma2}. Hence we consider also
a second double exponential disk,
\begin{equation}
\rho = \frac{\Sigma_0}{2 r_z}\exp\left({-\frac{z}{r_z}}\right)\exp\left({\frac{r_0-r}{r_d}}\right), 
\end{equation}
 whose parameters are not fixed. This component is allowed to serve as
lenses.  Analysis of the vertical velocity distributions of stars in the
vicinity of the sun, gives $2\sigma$ upper limits on the total surface
density within $1\kpc$ of the disk plane of at most $85\Msol
\pc2$\cite{sigma0,sigma1,sigma2}. Of this about $30\Msol \pc2$ is in the form of
bright stars and gas, while perhaps $10\Msol \pc2$ is in the form of M
dwarfs which could serve as lenses. About $10\Msol \pc2$ is
contributed by the halo, more in flattened models. This leaves a maximum
of $35\Msol \pc2$ for a possible dark disk. Adding in the M dwarf
lenses, the surface density of lenses should be in the range $10-45\Msol
\pc2$. To be conservative we choose the range $10-55\Msol \pc2$. The
scale height for the dark component of the disk is unknown. We therefore
consider the range $0.2-1.5\kpc$, which corresponds to populations tighter
than the visible stars at the low end, to a significantly heated
population at the high end. We fix the disk scale length at a value of 3.5kpc\cite{scalelength}.

\section{Data}

\subsection{Events \& Fields}
The basic requirement for microlensing searches is a very large sample of
distant stars. For studies of the halo the obvious choices of fields are
towards the Large Magellanic Cloud and the Small Magellanic
Cloud. Searches towards the bulge are complicated by the extinction from
intervening dust obscuring the Galactic Center. In the red and blue bands
used by the MACHO collaboration, there are only a relatively few fields
with a large enough density of stars. One of these ``holes in the dust''
is Baade's Window at Galactic longitude (l) and latitude (b) (1$\deg$,-4$\deg$),
towards which the first-year observations are clustered.  We show in
Fig.~1 a plot of the 24 fields reported on by the MACHO
collaboration. For reference, also included are some of the OGLE fields,
which are also of relatively high density.  The MACHO collaboration observed
12.6 million stars for a period of 190 days during the 1993 season in
these 24 fields\cite{MACHO1st}. The 41 events that pass their criteria
for microlensing are plotted on Fig.~1 as dots with radii proportional
to their duration. The location and duration of each event can be found in
Alcock et al. Table 1\cite{MACHO1st}. In addition to the location and duration of the
events we also need a variety of other numbers. Information on the
location, size and orientation of the 24 fields can be found at the MACHO
website, http://wwwmacho.anu.edu.au.
\begin{figure}[htb]
\epsfysize=13.5cm
\centerline{
} %\rotate[r]{\epsfbox{figure1.eps}}}
\caption[$20\deg \times 20\deg$ region centered at the
galactic center.]{$20\deg \times 20\deg$ region centered at the
galactic center. Light boxes are the 24 MACHO 1st year fields. Smaller
heavy boxes are OGLE fields. The 41 events we analyse are shown as dots
with radii proportional to the event duration.}
\end{figure}
The two quantities that pose the greatest problems are the observational
efficiencies and the density of the stars in each field. As of this
writing, the MACHO collaboration has not completed its analysis of the
full blending efficiencies for the bulge fields. As a reasonable
approximation, they suggest using their sampling efficiencies with a 0.75
factor correction. Accordingly, we following this prescription and use
their standard cut sampling efficiencies from Fig.~5 of Alcock et
al\cite{MACHO1st}. Since the fields under consideration are similar and all are crowding
limited, we assume that the efficiencies are uniform across the sample. The
reader is warned however, that this is a major source of uncertainty. It
is possible that the efficiency is not only a function of duration but
also of position \cite{HanGould} which would introduce spurious spatial
structure into the microlensing distribution.

	The final quantity we need for our analysis is the density of
observed sources. In view of the fact that the fields are crowding
limited and in the absence of better data, we assume a uniform density of
source stars across the fields. Accounting for overlap we get $\sigma=
1.06\times 10^6 {\rm deg}^{-2}$. We note an encouraging point. The
efficiencies are likely to get better as the fields get less crowded since
blending effects are smaller. On the other hand, less crowded fields mean a
smaller source density. Hence the errors due to our assumptions of
constant efficiencies and source densities should be in opposite
directions and are likely to at least partly cancel.

\subsection{Structure of the Data}
Before attempting to extract information about the structure of the Galaxy
from the data, we first look at what we can learn about the structure of the data itself.  
We apply our likelihood formalism to the very simplest model of the
optical depth possible:
\begin{equation}
\tau = \tau_0,
\end{equation}
a flat optical depth over the entire set of fields. We obtain the result
$\tau_0 = 1.93\pm0.39\times 10^{-6}$ where we quote 68\% confidence
limits. The MACHO collaboration's reported value of
$2.4\pm0.5\times10^{-6}$ for the same events includes a ``correction''
factor of $1/0.8$ introduced to adjust for contamination by source stars
in the disk. Undoing this ``correction'', we obtain
$1.9\pm0.4\times10^{-6}$, in agreement with our result.
Note that our errors are from our analytic calculations and not the
result of Monte Carlo simulations.

\subsection{Gradients}
We next apply our formalism to a slightly more complicated set of models:
those with an optical depth gradient in the $b$ and $l$ directions. We consider models with the form,
\begin{equation}
\tau = \tau_0 + \frac{d\tau}{db} (b-b_0) + \frac{d\tau}{dl} (l-l_0),
\end{equation}
where $b_0=-4$ and $l_0=2$. Fig~.2 shows contours of the likelihood,
marginalized over $\tau_0$, in the $\frac{d\tau}{db}$ --
$\frac{d\tau}{dl}$ plane. A gradient in the optical depth in latitude is
clearly indicated. The case for a slope in longitude is less
clear-cut. Marginalizing over the remaining parameter in each case we
obtain $\frac{d\tau}{db}=1.12\pm0.37\times10^{-6}/{\rm deg}$ and
$\frac{d\tau}{dl}=-1.71\pm1.19\times10^{-7}/{\rm deg}$, again with 68\%
confidence limits.

Our latitude gradient can be compared with an estimate based on the MACHO
clump giant optical depths reported for fields above and below
$b=-3.5\deg$\cite{MACHO1st}. Scaling the calculated slope,
$s=(6.32-1.57)\times10^{-6}/1.38\deg$, to the full sample and undoing
their $1/0.80$ disk correction, we obtain an estimate of
$\frac{d\tau}{db}=1.7\times10^{-6}/{\rm deg}$ with likely errors of at
least $1.0\times10^{-6}/{\rm deg}$. This is fully consistent with our
results.

\subsection{Maps}
	Just how much information about the structure of the event
distribution can we extract? 
Our likelihood method lends itself nicely to the construction of the
model independent ``most likely'' map of the microlensing optical
depth. Consider a general function $\tau(b,l)$ giving the optical
depth. We would like to find the positive definite function $\tau(b,l)$
which maximizes the likelihood or equivalently minimize the negative log likelihood, 
\begin{equation}
LL = \int\int \left[ \frac{T}{t_0}\tau(b,l) \sigma(b,l) - Q(b,l) \log{(\tau(b,l))} \right] db dl  
\end{equation}
where 
\BE Q=\frac{\pi}{4 t_0} \sum_{events}\hat{t}_i \delta(l-l_i,b-b_i).
\EE 
Taken alone, however, this condition is insufficient. The solutions to
this equation turn out to be delta functions at the event locations, a
clearly unphysical situation. What is missing is that the optical depth
should be a smooth function of position on the sky. Thus we must add a
smoothing term to the log likelihood to be minimized. Our smoothing term
must discourage structure unmotivated by the data, yet at the same time
not penalize legitimate gradients such as we have seen in the data. We
choose to minimize the extrinsic curvature, given by 
\BE 
K= \frac{d^2\tau}{db^2}+\frac{d^2\tau}{dl^2}+2\frac{d^2\tau}{dldb}.  
\EE 
Thus we require a minimum of
\begin{equation}
LL = \int\int \left[ \frac{T}{t_0} \tau(b,l) \sigma(b,l) - Q(b,l) \log{(\tau(b,l))} + \lambda K^2\right] db dl 
\end{equation}
where $\lambda$ controls how much smoothing we require.  

We solve for $\tau$ by an iterative scheme, starting with a flat
$\tau(l,b)$. Setting $\lambda$ to provide a reasonable smoothness we
produce the map shown in Fig.~3.  Our generated map contains few
surprises. We note a pronounced tilt in galactic latitude and a small one
in galactic longitude just as we saw in the earlier sections. A slight
bending of the contours to wrap around the Galactic center is also
present. It is important to remember that although the generated map is
smooth and does not appear ``noisy'', this is an artifact of the way it is
created: the smoothing term ensures that the resulting map is fairly
smooth. The significance of the present map is low, due to the small
number of events. A considerably larger data set would be
needed before the map could be used to yield detailed information.

\section{Results}
	With the results from our look at the data in mind, we now consider
the more realistic models of the Galaxy discussed above. For each type
of bulge, G2 and E2, we calculate the likelihood as a function of the
various Galactic parameters with the stated ranges. Since the functional
form of the bulge is so poorly constrained, we resist the
temptation to make a direct comparison between the two bulge models. Given
the number of parameters in our models and the present number of events,
any such comparison would be of marginal significance. Instead, we focus
on the parameters which have meaning independent of the functional form such
as the mass of the bar, $M_B$, the inclination angle of the bar,
$\theta_B$ and the bar axis ratio, $r=\frac{x_0}{y_0}$. In this section we
will discuss the limits we can put on bulges of only these functional
forms. 

\subsection{Bar}
	Due to the large number of parameters in our full model and limits
on computational power, in our exploration of the implication of the
microlensing events for bulge parameters we fix the disk parameters to
reasonable values: $\Sigma_0 = 30.0, r_z=0.3\kpc,$ and $r_d=3.5\kpc$,
while varying $M_B$, $\theta_B$, $x_0$ and $y_0$. The bulge quantities we
are most interested in, $M_B$, $\theta_B$ and $r$, should be most strongly
influenced by the magnitude and longitudinal gradient of the observed
optical depth. Of these, only the magnitude of the optical depth is
sensitive to the disk parameters. We expect that as we increase the disk
surface density, the inferred mass will decrease. We have checked our
results and find that this effect amounts to a less than
$0.3\times10^{10}\Msol$ shift even when we increase the disk to $55\Msol
\pc2.$ Our limits on other bulge parameters are unaffected.

\subsubsection{Bar Mass and Orientation}
Our major results concerning the bar mass and orientation are summarized
in Figs.~4\&5. These figures show contours of likelihood as a function of
the mass of the bar and orientation angle away from our line of sight. The
two horizontal scale lengths, $x_0$ and $y_0$ were marginalized. Fig.~4
shows results for G2 models, while Fig.~5 presents E2 models. We discuss
the G2 case first.

	One feature of Fig.~4 is immediately apparent: mass can be traded
off for angle. A ridge in the likelihood lies on the line
$M_{BAR}(10^{10}\Msol)-0.11\theta_B({\rm deg})=1.4$. There are limits to
this trade-off, however. If the mass of the bar is much beyond
$4.0\times10^{10}\Msol$, too high an optical depth will be produced to fit
the data even if the angle is increased dramatically. There is also a
sharp cutoff when the mass drops to below about 1.7. At such low masses,
decreasing the angle no longer helps but rather hurts since the maximum
optical depth is at a non-zero angle.\cite{lateZhao} The situation is much
the same for the E2 models (Fig.~5), except shifted by about
$1.1\times10^{10}\Msol$ in bulge mass. The E2 models drop off too rapidly
to be efficient at producing microlensing optical depth even when
optimally aligned, and hence need considerably higher bar
masses\cite{lateZhao}. Below a mass of about $2.0\times10^{10}\Msol$ it
becomes difficult to produce the high optical depths required by the
data. We show in Fig.~6 the likelihood for the bar mass now marginalized
over the bar orientation as well. The 90\% confidence limit is at
$1.75\times10^{10}\Msol$ for the G2 bulge and $2.6\times10^{10}\Msol$ for
the E2 bulge. Taking into account the uncertainty in the disk
normalization, we arrive at lower bounds on the bulge mass of
$1.5\times10^{10}\Msol$ (G2) and $2.3\times10^{10}\Msol$ (E2). The most
likely values are around $3.5\times10^{10}\Msol$ for G2 models and beyond
$4.0\times10^{10}\Msol$ for E2 models.

 In Fig.~7 we plot the marginalized likelihood versus bulge inclination
angle. Low inclination angles are clearly favored. The likelihood has
dropped off strongly by $30\deg$ and $20\deg$ for the G2 and E2 models
respectively. Beyond these angles, even bar masses as high as
$4.0\times10^{10}\Msol$ cannot produce enough microlensing to be
compatible with the experimental results. The $90\%$ confidence
limits are $30\deg$ (G2) and $21\deg$ (E2).

Our results, using only a single year of microlensing data, fit very
nicely with attempts to constraint the bulge mass and orientation angle by
other means. Analysis of the stellar motions in the bulge gives a range of
mass estimates from slightly below $2.0\times10^{10}\Msol$ to almost
$3.0\times10^{10}\Msol$.\cite{Kent,zhao,blum} Several authors have derived
bulge masses around $2.2\times10^{10}\Msol$ using a variety of methods
based on modeling of the gas content of the bulge.\cite{Binney,ZRS96} A
simple argument by Zhao et al. gives this as an upper
limit\cite{lateZhao}. Our limit on the bulge mass is consistent with the
reported values for the bulge mass.

The OGLE collaboration has reported an analysis of the distribution of the
bulge red clump giants within their fields shown in Fig.~1. They report
a bulge orientation of between 20 and $30\deg$ almost independent of the
bulge model.\cite{OGLEfields} Dwek et al. analyze the DIRBE maps of the
infrared emmision from the bulge and conclude that the orientation angle
lies in the range $10-40\deg$. Our range of $0-30\deg$ is also consistent
with the value $16\pm2\deg$ suggested by Binney et al.'s analysis of gas
dynamics\cite{Binney}.  

Perhaps more important than our limits on $M_B$ and $\theta_B$ separately,
are the full contours of likelihood in the $M_B$ - $\theta_B$ plane
showing the correlation between high bulge mass and high orientation
angle.  As we discuss later, increases in the number of events at Baade's
window do little to break the $M_B$ - $\theta_B$ degeneracy, but make the
ridge considerably narrower.  It requires more information to uniquely
pick out a bulge mass. An analysis using the tensor virial theorem for the
bulge by Blum \cite{blum} gives exactly the opposite degeneracy: high mass
is correlated with low angle. Thus the results of microlensing and
dynamics arguments are complementary and may be able to break the
mass-angle degeneracy for either alone.

\subsubsection{Bar Axis Ratios}
	The other bulge quantity we look at is the bar axis ratio
$r=\frac{x_0}{y_0}$. Although we do not explicitly have the axis ratio as an
input quantity for our bulge models, microlensing puts limits on the ratio
of the scale lengths. In Figs.~8\&9 we have marginalized $\theta_B$ and
the $x_0$-$y_0$ pair subject to the constraint $r=\frac{x_0}{y_0}$.  One
can immediately see a trade-off between the axis ratio and the mass. A
higher axis ratio concentrates the mass where it will do the most
microlensing. Hence, in conjunction with a low orientation angle, this
allows a lower mass. Since a bar-like configuration seems to be favored by
the most recent data on the bulge it is interesting to note that our
results do not completely rule out axisymmetric models. The preferred
axisymmetric models have very high masses. On the other hand, measurements of velocity
dispersions in the bulge give low constraints on masses of axisymmetric
models.  The strongest microlensing evidence against
axisymmetric models comes from a consideration of the more limited bulge
giant subsample of the events, which probe the optical depth for sources
distributed throughout the bar. An axisymmetric model can not produce
enough microlensing to account for the high optical depth implied by these
events\cite{long}. Our most likely values, r=3.5 (G2) and r=2.5 (E2), are
consistent with the axis ratios determined by other means. Dwek et
al. give a range $2.5-5.0$ for their models\cite{Dwek}. Stanek et
al. obtain a value in the range $2.0-2.5$ again independent of model
choice\cite{OGLEfields}. The distribution of bulge Mira variables gives an
axis ratio of 3.9.\cite{whitelock}.

\subsection{Disk-Bulge Discrimination}
Since we hope to discriminate between the bulge and disk on the basis of
the latitude gradient information, the parameters $M_B$, $\theta_B$,
$r_z$, and $\Sigma_0$ are most relevant.  Thus we vary these quantities
while holding the two horizontal bulge scale lengths constant at
$x_0=1.58\kpc$ and $y_0=0.62\kpc$ for G2 models. The results for the E2
models are similar to those for the G2 models, except for a shift in the
bulge mass as noted above. Since we are interested only in the disk
parameters we discuss only the results for the G2 models.

\subsubsection{Surface Density}
	Fig.~10 shows the marginalized likelihood as a function of the
surface density of the disk versus mass of the bulge. It is immediately
clear that we cannot uniquely fix the surface density of the disk. Rather,
as was the case with the orientation angle of the bulge, $\Sigma_0$ shows
a linear degeneracy with $M_{BAR}$, with the ridge of the likelihood at
$M_{B}(10^{10}\Msol)+0.0088\Sigma_0(\Msol/pc^2)= 2.88.$ A heavy disk can
add as much as $0.7\times10^{-6}$ to the optical depth, allowing the bar
to be less massive. The widening of the likelihood contours towards the
top shows a slight tendency towards a more massive disk. This tendency,
although not significant, is due to the slope of the optical depth in
latitude.

\subsubsection{Scale Height}
	We show in Fig.~11 the likelihood as a function of the scale
height of the disk and the bar mass. The scale height is only very weakly
correlated with the bar mass and is basically unconstrained. There is a
spreading of the likelihood contours for low scale heights possibly
favoring scale heights below about $0.6\kpc$. At low scale lengths the
gradient in optical depth of the disk contribution is high. As the scale
length increases, the disk gradient decreases. Hence low values of $r_z$
are favored, with high values, where the gradient is low, suppressed.

\subsubsection{Disk Bulge Degeneracy}
	Our results for disk parameters are something of a
disappointment. With the present data we can say virtually nothing about
the disk structure. Part of the problem is the trade-off that occurs
between $M_B$ and $\Sigma_0$, keeping the optical depth constant and
producing the ridge in the $M_B$ versus $\Sigma_0$ likelihood plot. We had
hoped, however, that gradient information would allow us to break the
degeneracy between $M_B$ and $\Sigma_0$. The reason that this does not
happen is not that the area that the MACHO fields span is too small to
show strong structure with these few events; slopes in $b$ and $l$ are
indicated. The problem lies with the position of Baade's window and the
details of the Galactic models. Firstly, because the longitudes of the bulk
of the MACHO fields are low, the longitude slope expected in this region
is small for virtually any model. This makes the slope in longitude a poor
diagnostic. Secondly, at the location of Baade's window the $M_B$ -
$\Sigma_0$ trade-off also keeps the latitude gradient fairly constant over
a large range. We show this in Fig.~12, where the latitude gradient is
shown as a function of $\Sigma_0$, with the optical depth kept constant by
varying the $M_B$. The different curves show various values of the scale
height of the disk. Since the $b$ slope in the region of Baade's window is
constant for a wide range of models it is very difficult to discriminate
among them. This is why we see only hints of structure in our likelihood
plots.

\section{Future Directions}
	Over the next year, two collaborations, OGLE and EROS, will be
moving to dedicated telescopes and fully automated analysis
systems. Together with the currently running MACHO system, these groups
have the potential for producing many times the data we have used
here. How will such a wealth of data effect our results? To explore this
question we have synthesized 4 years of observations and rerun our
analysis. Our synthetic observations were constructed assuming a G2 model
with $M_B=3.0\times10^{10}\Msol, x_0=1.58\kpc, y_0=0.62\kpc,
\theta_B=15\deg, \Sigma_0=30.0\Msol \pc2, r_z=0.3/kpc$. The total
number of events expected was calculated as \BE N_{exp}= \frac{T}{\langle
t \rangle}\int \sigma\tau(l,b) dl db ~~~~~~~~{\rm w}/ \langle t
\rangle=\frac{\pi}{4}\sum_i\frac{\hat{t}_i}{\epsilon_i}/\sum_i\frac{1}{\epsilon_i},
\EE where the average was done over the present data set and the integral
is over the MACHO first-year fields. From this expected number, the actual
number was picked assuming Poisson statistics. Since the rate is
proportional to the optical depth (assuming constant average duration),
the events were laid down randomly according to the optical depth. A
duration was picked out of the efficiency weighted set of observed
durations. The proposed event was then ``observed'' or not based on the
efficiency for that duration. The final set of ``observed'' events was run
through our analysis. The results are shown in Fig.~13.  We note a
number of features of the results. First is that the confidence regions
have tightened somewhat as expected. Second, the basic degeneracy between
$M_B$ and $\theta_B$ is uneffected. Although $\theta_B$
now seems to be quite well constrained, $M_B$ still varies over a wide
range. As we discussed earlier, part of the problem is in the distribution
of our present lines of sight. Baade's window is a poor location for
determining longitude slope, and the latitude slope is correlated with the
optical depth, making it less useful as a diagnostic tool. If we wish to
determine the bulge parameters solely from microlensing, simply collecting
more data in the same fields will not easily break the degeneracies
between parameters.  We must look to expanding our range of fields.

	So, what is the best strategy to use? Microlensing searches are
costly and very time-consuming: we would like to find a strategy that
maximizes the scientific return. We attempt to address this issue by
analyzing a set of very stylized strategies that will allow us some
insight into real searches.
	
	Let us assume that we observe four identical fields centered at
$(1.0,-3.0)$, $(1.0+\Delta l,-3.0)$, $(1.0,-3.0-\Delta b)$ and $(1.0+\Delta
l,-3.0-\Delta b)$.  Further, we take the limit as the size of the
fields goes to zero but the exposure for each field, $E$, stays constant. In
this way the only gradient information will come from the field separation,
and not from the distribution of events within the fields. Each choice of $\Delta
l$, $\Delta b$ will be a distinct strategy. Let $\hat{t}_i^j$ represent
the durations of the events observed in the $j$-th field. Then the likelihood
for a given model, $\tau(l,b)$ is
\BE
L=\prod_j e^{-\frac{E}{t_0}(\tau_j)}(\sigma\tau_j)^{\frac{\pi}{4t_0}\sum_i\frac{\hat{t}_i^j}{\epsilon_i}}
\EE
where $\tau_j$ is the predicted optical depth at the $j$-th field.
Let $\tau^0(l,b)$ represent the optical depth of the underlying
model. Then on average
$\frac{\pi}{4t_0}\sum_i\frac{\hat{t}_i^j}{\epsilon_i}=\frac{E}{t_0}\tau^0_j$
and we will have
\BE
L=\prod_j e^{-\frac{E}{t_0}(\tau_j)}(\sigma\tau_j)^{\frac{E}{t_0}\tau^0_j}.
\EE
Using this likelihood we can now determine how well a given strategy can
recover our underlying model. Figs.~14,15\&16 show the magnitude of the 68\%
confidence intervals in various quantities as a function of the longitude
and latitude separation of the fields assuming twice the exposure of
present experiments. We used the same model as in the previous section. The first thing one notices is
that the varying strategies don't make as much difference as might be
hoped. As the separation of fields is increased, our lever arm for making
determinations of the quantities increases. However, at the same time the
outer fields are in regions with low $\tau$ and hence few events and poor
statistics. These two effects tend to cancel out, making dramatic
improvements difficult. Nevertheless, what can we learn from these graphs?
First of all, the present data correspond to roughly $\Delta l=3.0$,
$\Delta b=2.0$. This is very close to the worst possible region for
determination of all three parameters.

	If we wish to improve our determination of $M_B$ the results
suggest a relatively large $\Delta b$ and a smallish $\Delta l$. For
$\Sigma_0$, on the other hand, we should have a high $\Delta l$ and $\Delta
b$ is irrelevant. Finally, for $\theta_B$, a moderate $\Delta l$ is
required and again $\Delta b$ is mostly unimportant. A single search
strategy to fix the parameters would need to probe the entire range of
scales in longitude. Coverage in latitude appears to be less important;
only the large scales need to be probed. The best strategy would seem to
be one which includes many fields scattered over the entire bulge instead
of concentrated in one region. Although the optical depth at any given
location would be less well defined, better limits on the global
parameters would be obtained.

\section{Conclusions}
In this paper we have developed a novel likelihood technique for the
analysis of the microlensing data towards the bulge. We construct a
likelihood that is sensitive to the spatial distribution of the events.
Our technique is both more flexible than calculations that have been done
before, and allows for a direct comparison of the data to models of the
mass distribution for the Galaxy. It is particularly good for dealing with
data from more than one line of sight, field overlaps and variations in
the density of stars observed. Its sensitivity to the position of events
makes it ideal for determining gradients in the optical depth.  Applying
this technique to the first year MACHO data we have confirmed the strong
slope in latitude found by the MACHO collaboration, and found hints of
one in longitude. We have for the first time, given quantitative measure
of these slopes, $\frac{d\tau}{dl}=-1.71\pm1.19\times10^{-7}/{\rm deg}$
and $\frac{d\tau}{db}=1.12\pm0.37\times10^{-6}/{\rm deg}$. We also apply
our analysis to constructing a crude ``most likely'' map of the
microlensing optical depth over the observed region.

	We confront a set of Galactic models consisting of a bulge with
either G2 or E2 functional form and an exponential disk, with the
data. With only one season of microlensing data, we can already set
meaningful limits on various bulge parameters. We find that $M_B>1.5
(2.0)\times10^{10}\Msol$, for the G2 (E2) based models. Most likely values
for the bulge mass are much higher: $M_B=3.5 (>4.0)\times10^{10}\Msol$.
Previous work has shown that such high bulge masses imply low halo MACHO
fractions\cite{long}.  A massive bulge puts tight constraints on the
contribution of the disk to the rotation curve at small radii. A small
disk, however, leaves more room for the halo in the outer rotation curve,
implying a massive halo. Since microlensing results towards the LMC fix
the MACHO content of the halo, a massive halo implies a smaller MACHO
fraction.

We also constrain the inclination angle of the bulge finding that
$\theta_B < 30\deg (21\deg)$, consistent with other measurements. Our most
likely values for the axis ratio of the bulge, $r=\frac{x_0}{y_0} = 3.5
(2.5)$, are consistent with determinations by other methods. We note that
axisymmetric bulge models are not entirely ruled out with the full sample
of events. Such models, however, typically need very high
($\approx4.0\times10^{10}\Msol)$ bulge masses. Such high bulge masses are
unlikely for axisymmetric bulge models\cite{kentpriv}. No limits could be set on
the disk component due to a degeneracy in the latitude slope of the
optical depth between the bulge and disk contributions. We discuss what
can be expected with an increase in the number of seasons of data.

	Finally, we have attempted to quantitatively discuss various
strategies for microlensing searches and conclude that a strategy
observing many fields well scattered in longitude offers the best return
in determining bulge and disk parameters. The distribution of fields in
latitude is less important. Despite the difficulties, this is perhaps more
important than simply getting the optical depth more accurately at one
location as it will allow a better determination of the relative
contributions of the bar and the disk. Ideally, the optical depth can be
mapped over a wide range in both latitude and longitude, yielding detailed
information about the mass distribution in the inner Galaxy. The field of
microlensing promises to be an eventful one for the foreseeable future!

\section*{Acknowledgments}
This work was supported by the DOE (at Chicago and Fermilab) and the NASA
(at Fermilab through grant NAG 5-2788).  I would like to thank Craig
J. Copi for many stimulating conversations, and for his gracious work on
the hMgo package. I would also like to thank Michael Turner, Evalyn Gates
and Robert Nichol.

\section*{Figure Captions}
\bigskip

\noindent{\bf Figure 1:} {$20\deg \times 20\deg$ region centered at the
galactic center. Light boxes are the 24 MACHO 1st year fields. Smaller
heavy boxes are OGLE fields. The 41 events we analyse are shown as dots
with radii proportional to the event duration.}

\medskip
\noindent{\bf Figure 2:} {Likelihood contours for $l$ and $b$ gradients in
the optical depth. Solid lines are the 68\% confidence contours. Dotted
lines denote 95\% confidence. Dashed lines denote 99\% confidence. Also
included, to guide the eye are long dashed lines for 38\% confidence.}

\medskip
\noindent{\bf Figure 3:} {``Most likely'' map of the optical depth based on
the first year events. The solid lines show contours of $4.0, 3.0, 2.0,
1.0, {\rm and} 0.5\times10^{-6}$ from the top down.} 

\medskip
\noindent{\bf Figure 4:} {Contours of likelihood in the $\theta_B - M_B$
plane for G2 models with $x_0$ and $y_0$ marginalized. Disk values where
held fixed at $\Sigma_0=30\Msol pc^-2$ and $r_z=0.3\kpc$. Contours are as
in Fig.~2.}

\medskip
\noindent{\bf Figure 5:} {Same as Fig.~4. but for E2 models.}

\medskip
\noindent{\bf Figure 6:} {Likelihoods from Figs.~4\&5 now with $\theta_B$
marginalized. Solid line for G2 models, dashed line for E2 models.}

\medskip
\noindent{\bf Figure 7:} {Same as Fig.~6, but with $M_B$ marginalized.}

\medskip
\noindent{\bf Figure 8:} {Contours of likelihood in the $M_B$ - $r$(axis
ratio) plane. $\theta_B$ and one degree of freedom from $x_0$, $y_0$ have
been marginalized. As in Fig.~4, disk parameters have been
fixed. Contours are as in Fig.~2.}

\medskip
\noindent{\bf Figure 9:} {Same as Fig.~8 but for E2 models.}

\medskip
\noindent{\bf Figure 10:} {Contours of likelihood in the $M_B$ - $\Sigma_0$
plane. $\theta_B$ and $r_z$ have been marginalized. Bulge parameters $x_0$
and $y_0$ have been held fixed at $x_0=1.58\kpc$ and
$y_0=0.62\kpc$. Contours are as in Fig.~2.}

\medskip
\noindent{\bf Figure 11:} {Contours of likelihood in the $M_B$ - $r_z$
plane. $\theta_B$ and $\Sigma_0$ have been marginalized. Bulge parameters $x_0$
and $y_0$ have been held fixed at $x_0=1.58\kpc$ and
$y_0=0.62\kpc$. Contours are as in Fig.~2.}

\medskip
\noindent{\bf Figure 12:} {Slope in latitude for the optical depth as a
function of $\Sigma_0$, the disk surface density. The total optical depth
has been held constant by varying $M_B$ as $\Sigma_0$ varies. The lines
represent: solid ($r_z=0.3\kpc$), dashed ($r_z=0.5\kpc$), long dashed
($r_z=1.0\kpc$), and  dot-dashed ($r_z=1.5\kpc$). Total variation across our
range of disk surface densities is less than 10\%. }

\medskip
\noindent{\bf Figure 13:} {Likelihood in the $M_B$ - $\theta_B$ plane
for a simulated 4 seasons of data based on a G2 model with
$M_B=3.0\times10^{10}\Msol, x_0=1.58\kpc, y_0=0.62\kpc, \theta_B=15\deg,
\Sigma_0=30.0\Msol \pc2, r_z=0.3/kpc$. Contours same as Fig.~2. }

\medskip
\noindent{\bf Figure 14:} {Contours of the size of the 68\% confidence
regions in the determination of $M_B$. Values are calculated as a function
of the latitude and longitude spacing of the observed fields. Solid: 1.46,
Dotted: 1.53, Dashed: 1.61, Long Dash: 1.69$\times10^{10}\Msol$ }

\medskip
\noindent{\bf Figure 15:} {Same
as Fig.~14 but for $\theta_B$. Solid: 16$\deg$, Dotted: 19$\deg$, Dashed:
23$\deg$, Long Dash: 26$\deg$}

\medskip
\noindent{\bf Figure 16:} {Same as Fig.~14, but for $\Sigma_0$. Solid:
29, Dotted: 35, Dashed: 42, Long Dash: 48$\Msol \pc2$.}

\end{document}